\documentclass[aps, prb, twocolumn, floatfix]{revtex4-2}

\usepackage[dvips]{graphicx}
\usepackage{epsfig}

\usepackage[margin=2cm,head=0.5cm]{geometry}
\usepackage{bm}
\usepackage{amsmath}
\usepackage{amssymb}
\usepackage{latexsym}
\usepackage{amsfonts}
\usepackage{epsfig}
\usepackage{color}
\usepackage[linktocpage, colorlinks=true ,linkcolor=blue, citecolor=blue]{hyperref}
\usepackage[all]{hypcap}
\usepackage[utf8]{inputenc}
\usepackage{natbib} 

\newcommand{\expval}[1]{\left< #1 \right>}

\newcommand{\ket}[1]{\left|#1\right>}

\newcommand{\nn}{\nonumber\\}

\newcommand{\f}[1]{\mbox{\boldmath$#1$}}

\newcommand{\bea}{\begin{eqnarray}}
\newcommand{\eea}{\end{eqnarray}}
\newcommand{\beann}{\begin{eqnarray*}}
\newcommand{\eeann}{\end{eqnarray*}}

\newcommand{\abs}[1]{{\left| #1 \right|}}

\newcommand{\asinh}{\rm arsinh}

\newcommand{\ii}{\mathrm{i}}  

\begin{document}
  
\title{Transport statistics and thermoelectric performance of Hofstadters butterfly}

\author{F. Quei{\ss}er}
\author{G. Schaller}
\email{g.schaller@hzdr.de}
\affiliation{Helmholtz-Zentrum Dresden-Rossendorf, Bautzner Landstraße 400, 01328 Dresden, Germany}

\begin{abstract}
We investigate electrons subject to a magnetic field that may tunnel along a square lattice and additionally to two electronic leads.
Using non-equilibrium Greens functions we evaluate the transmission and from that particle and energy currents and zero-frequency noise
of the setup. 
We identify sweet spots for which the transmission reaches unity over broad frequency windows -- carried by topologically protected boundary modes that are quite robust against
local fluctuations of on-site energies. 
These transmission plateaus can be exploited for the implementation robust current standards, clocks and thermoelectric devices.
\end{abstract}

\maketitle


\section{Introduction}

The discovery of the quantum Hall effect~\cite{vonklitzing1980a,laughlin1981a} established that electronic transport can be governed by the topology of the underlying band structure, giving rise to robust edge states and quantized transport coefficients.
This insight has motivated extensive research into topological materials~\cite{avron2003a,hastings2008b,yan2012a,schmidt2013a,ren2013b,riwar2016a,vergniory2019a,liu2019a} and their potential for applications ranging from robust electronic~\cite{niklas2016a,ruocco2017a,jin2023a} and photonic transport~\cite{ozawa2019a,khanikaev2024a} to robust qubits in quantum information processing~\cite{flensberg2021a}, sensing~\cite{sarkar2022a}, and energy conversion~\cite{xu2017a,mamede2023a,yang2025b}.

The presence of topologically protected boundary states provides a route toward transport that is remarkably insensitive to certain forms of disorder and imperfections. 
Clearly, these properties are of great interest for thermoelectric applications~\cite{xu2014a,sothmann2014a,xu2017a,hajiloo2020a,myers2022a,toriyama2024a,kajola2025a}, where the interplay between electrical and thermal transport can be exploited to control energy and heat flows.

The Hofstadter model~\cite{hofstadter1976a} provides a paradigmatic setting in which these topological transport phenomena can be studied. 
While the resulting spectral properties have been extensively investigated theoretically and experimentally~\cite{janecek2013a,naumis2016a,wackerl2019a,nuckolls2025a}, the statistical properties of transport through finite Hofstadter systems, particularly in the thermoelectric regime, remain comparatively less explored.
To provide a complementary perspective on topological systems that goes beyond boundary-driven 1d models~\cite{boehling2018a,xue2025a} (where thermoelectric functions may operate near Carnot efficiencies but at an extremely low overall output), further beyond near equilibrium studies~\cite{cortes2026a}, and also beyond the weak-coupling regime~\cite{rivas2017b}, in this paper we analyze 
the first two transport cumulants of electronic transport through a finite-size Hof\-stadter model using non-equilibrium Greens functions.
We introduce the model below in Sec.~\ref{SEC:model}, our methods in Sec.~\ref{SEC:methods}, and afterwards present our results on the transmission, the bond currents, the quality of the counting statistics and the thermoelectric performance in Sec.~\ref{SEC:results}.
After concluding with Sec.~\ref{SEC:summary}, we provide some additional information in the appendix.

\section{Model}\label{SEC:model}

\begin{figure}[ht]
\includegraphics[width=0.45\textwidth,clip=true]{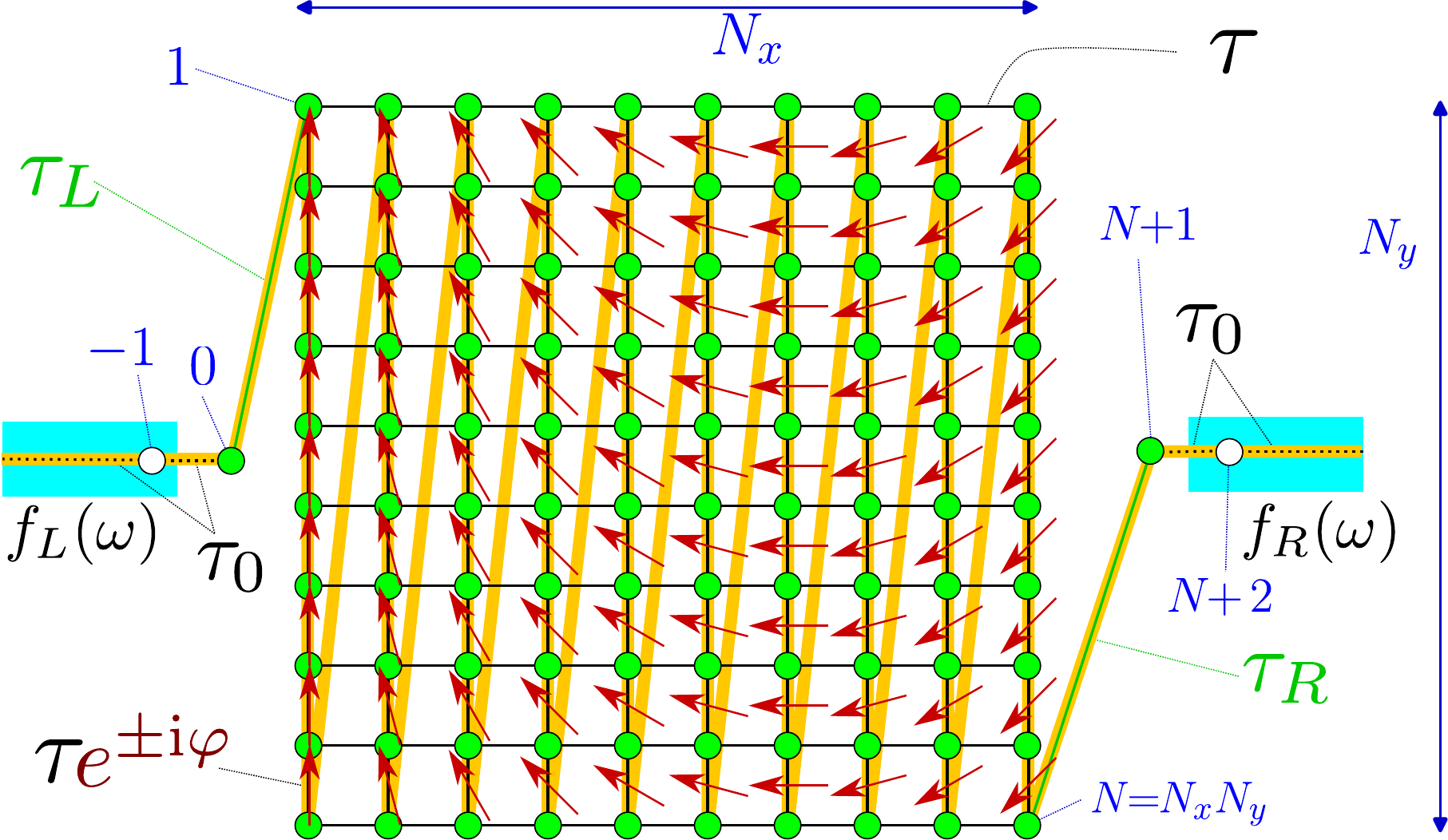}
\caption{\label{FIG:model_sketch}
Sketch of the model for $N_x=N_y=10$.
The left (sites $-\infty \ldots 0$) and right (sites $N+1 \ldots \infty$) semi-infinite chain reservoirs are described by local thermal equilibrium Fermi functions $f_{\nu}(\omega)$ and
bare tunneling amplitude $\tau_0$ (dotted black).
Inside the system (sites $1 \ldots N$), tunneling amplitudes (magnitude $\tau$, black lines) in $y$-direction have phase factors that depend on the horizontal position (red arrows).
Transport is enabled by coupling the system to the leads via amplitudes $\tau_\nu$ (green lines). 
The system together with the adjacent reservoir sites forms the central system (green circles).
The bold yellow curve in the background denotes the  ordering of the lattice sites.
}
\end{figure}
Our model (see Fig.~\ref{FIG:model_sketch} for an illustration) 
is composed of semi-infinite left and right leads $H_B^{(\nu)}$ with $\nu \in \{L,R\}$ that act as reservoirs, which via the tunnelling interaction 
$H_I^{(\nu)}$ can exchange particles with a central system $H_S$.
The total Hamiltonian is given by
\begin{align}
H = \sum_{\nu \in \{L, R\}} [H_B^{(\nu)} + H_I^{(\nu)}] + H_S\,.
\end{align}

Specifically, we model the reservoirs as 
\begin{align}
H_B^{(L)} &= \tau_0 \sum_{n=-\infty}^{-1} (c_n^\dagger c_{n+1} + c_{n+1}^\dagger c_n)\,,\nn
H_B^{(R)} &= \tau_0 \sum_{n=N+1}^{\infty} (c_n^\dagger c_{n+1} + c_{n+1}^\dagger c_n)\,,
\end{align}
where $c_n$ is a fermionic annihilation operator at site $n$ and the tunneling amplitude $\tau_0$ determines the width of the resulting semi-circular spectral function. 
Such leads can be extended in a straightforward way towards non-trivial spectral functions by additional non-homogeneous sites near the system~\cite{ehrlich2021a}.

At their ends, the leads are tunnel-coupled with amplitude $\tau_\nu$ to the system (composed of $N$ sites), such that the interaction Hamiltonians only concern two sites each
\begin{align}
H_I^{(L)} &= \tau_L (c_0^\dagger c_1 + c_1^\dagger c_0)\,,\nn
H_I^{(R)} &= \tau_R (c_N^\dagger c_{N+1} + c_{N+1}^\dagger c_N)\,.
\end{align}
Our model can be generalized to couple to the system at multiple sites -- which however would also raise questions on the magnetic flux at the contact area.
As the reservoirs are modeled by semi-infinite chains, this situation is equivalent to the problem of an inhomogeneous magnetic field inside the system and a single-site coupling (by shifting the system-reservoir boundary) and will be deferred to future work. 

Finally, the system is given by electrons on a square lattice with spacing $\Delta$, homogeneous tunneling amplitudes $\tau$ and subject to a perpendicular magnetic field $\vec B = B \vec e_z$.
Following the Peierls substitution~\cite{peierls1933a} $\tau_{R'\to R} \to \tau_{R'\to R} \exp\left\{\ii \frac{e}{\hbar} \int_{R'}^R \vec A(\vec r) \cdot d\vec r\right\}$ for a vector potential $\vec A = B x \vec e_y$, we arrive at the Hofstadter Hamiltonian~\cite{hofstadter1976a}
\begin{align}\label{EQ:ham_hofstadter}
H_S &= \tau \sum_{a=1}^{N_x-1} \sum_{b=1}^{N_y} 
\left[c_{a,b}^\dagger c_{a+1,b} + c_{a+1,b}^\dagger c_{a,b}\right]\nn
&\qquad+ \tau \sum_{a=1}^{N_x} \sum_{b=1}^{N_y-1}
\left[e^{+\ii {\cal B} a} c_{a,b}^\dagger c_{a,b+1} 
+ {\rm h.c.}\right]\,, 
\end{align}
where $\tau$ denotes the tunneling amplitude through the system and ${\cal B} = \frac{e}{\hbar} B \Delta^2$ denotes the magnetic flux per plaquette of area $\Delta^2$. 
Additionally, we labeled the sites of the central system with two indices via $n = (a-1)N_y+b$ with $N=N_x N_y$.
This Hamiltonian has trivial hopping amplitudes in $x$-direction and phase-modified hopping amplitudes in $y$-direction.
It is gauge-invariant with respect to changes of the vector potential that do not change the magnetic field $\vec{A} \to \vec{A} + \vec \nabla \Phi$:
Taking an arbitrary discretized gauge function on the 2d lattice $\Phi_{ab} = \Phi(a \Delta \vec{e}_x + b \Delta \vec{e}_y,t)$ (that in general could also be time-dependent), one would obtain a topologically equivalent Hamiltonian that is related to the above by the unitary rotation $c_{ab} \to c_{ab} e^{-\ii \Phi_{ab} e/\hbar}$ and thus has the very same spectrum.
Additionally, the Hofstadter Hamiltonian has the chiral symmetry $\Gamma H_S \Gamma = -H_S$ with chiral symmetry operator
$\Gamma = \exp\left\{\ii \pi \sum\limits_{ab:a+b={\rm even}} c_{ab}^\dagger c_{ab}\right\}$ obeying $\Gamma^2=\f{1}$,
which shows that its spectrum must be symmetric around zero: For every eigenvalue $\lambda$ and eigenstate $\ket{v}$ of $H_S$ one has that
$\Gamma \ket{v}$ is also an eigenstate of $H_S$ with eigenvalue $-\lambda$.

\begin{figure}[ht]
\begin{minipage}{0.8\linewidth}
\includegraphics[width=0.99\textwidth,clip=true]{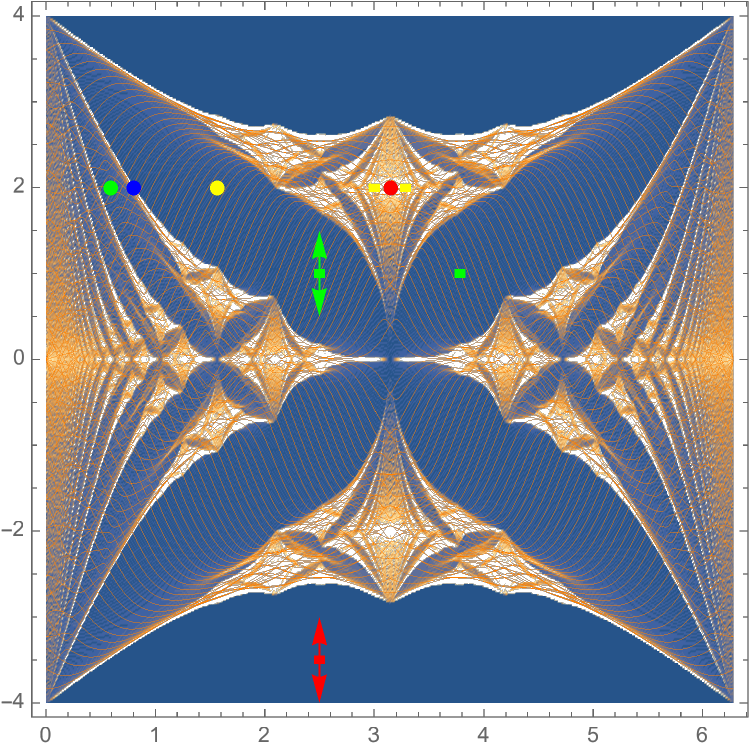}
\end{minipage}%
\begin{minipage}{0.15\linewidth}
\includegraphics[width=0.99\textwidth,clip=true]{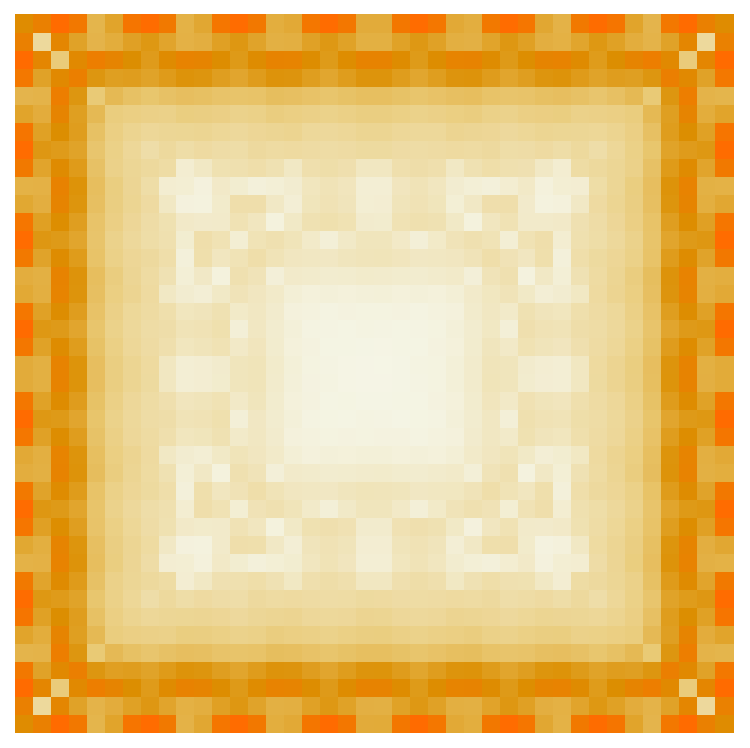}\\
\includegraphics[width=0.99\textwidth,clip=true]{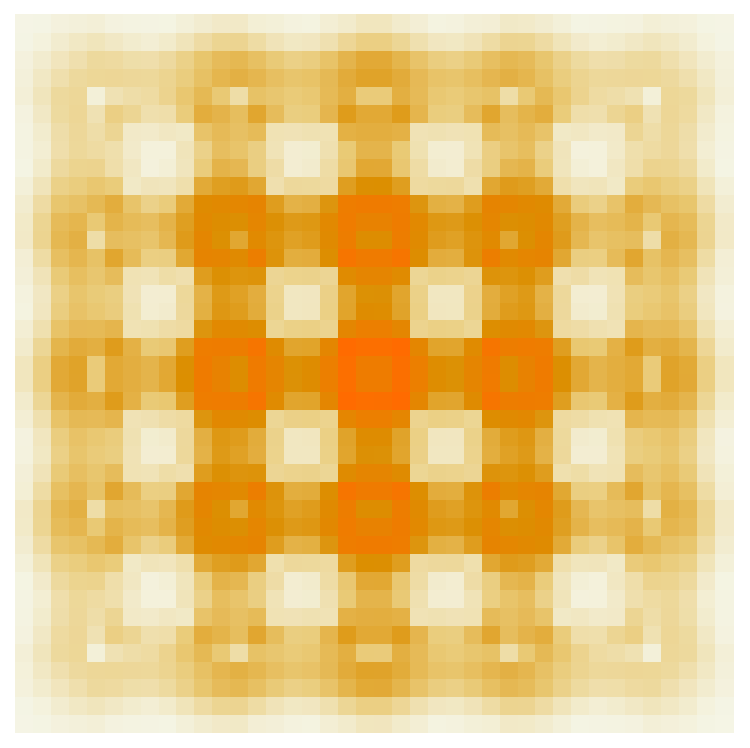}\\
\includegraphics[width=0.99\textwidth,clip=true]{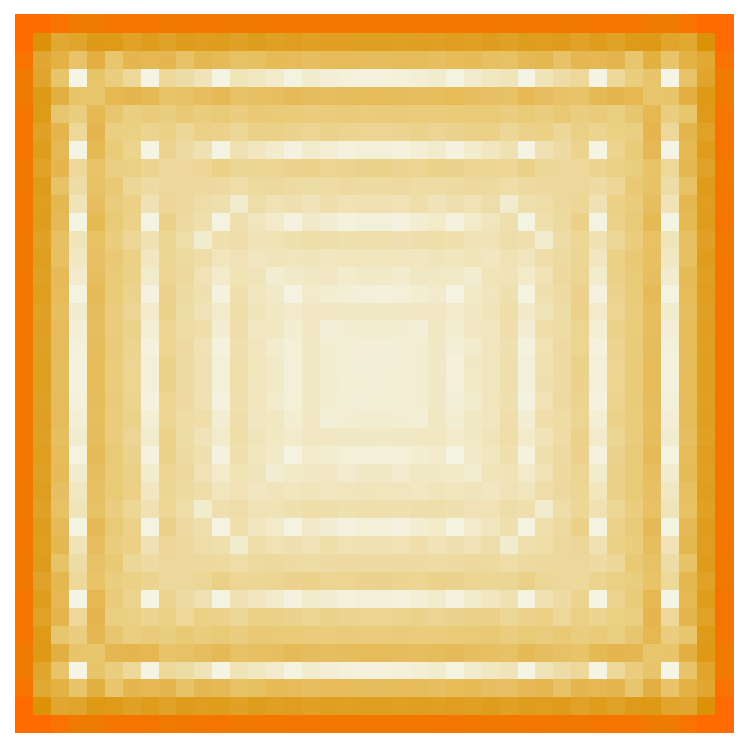}\\
\includegraphics[width=0.99\textwidth,clip=true]{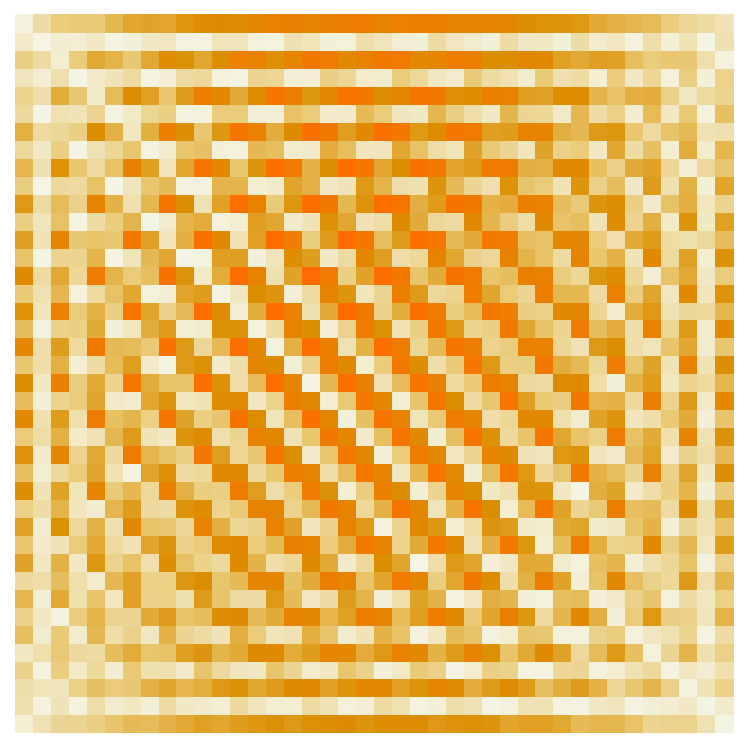}
\end{minipage}
\caption{\label{FIG:butterfly}
Level density of the Hofstadter Hamiltonian versus the magnetic field parameter ${\cal B} = \frac{e}{\hbar} B \Delta^2$ (horizontal axis) 
and energy (vertical axis, in units of $\tau$) for $N_x=N_y=40$ (background)
and actual spectrum for a $10\times 10$ grid (orange curves).
The characteristic butterfly structure is clearly visible, in the infinite-size limit $N_x,N_y\to\infty$ it develops a fractal
structure.
The blue regions inside the butterfly wings denote regions of lower level density populated only by boundary states.
Circles (from left to right) mark positions for which the amplitude distribution of exemplary eigenstates is shown
on the right (from top to bottom).
Within the butterfly wings (green and yellow dots), they correspond to surface modes, whereas in regions of high level density (blue and red dots), the wave function is distributed over the system.
Other symbols serve for orientation with other plots.
}
\end{figure}

Beyond the trivial values of ${\cal B}=0$ or ${\cal B}=\pi$ (see App.~\ref{APP:diagonalization}), analytic approaches to the eigenvalues are possible at specific points and require periodic boundary conditions~\cite{tesfaye2026a}.
A plot of the level density of the Hofstadter Hamiltonian for finite lattices together with exemplary wave functions 
is provided in Fig.~\ref{FIG:butterfly}.
One finds the typical Hofstadter butterfly spectrum with fractal structure.
In contrast to infinite lattices though, the butterfly wings are not empty but contain surface modes~\cite{analytis2004a} that can carry the transport.
We also mention here that experimental implementations of the Hofstadter Hamiltonian and related models that allow for sufficiently large magnetic flux have been reported in optical lattices~\cite{aidelsburger2013a,miyake2013a}.

\section{Methods}\label{SEC:methods}

\subsection{Transport characteristics}

Denoting the particle number operator of the right reservoir by $N_B^{(R)} = \sum_{n=N+1}^\infty c_n^\dagger c_n$, a central result of the Levitov-Lesovik formalism~\cite{levitov1993a} for non-interacting electronic transport is that  the stationary matter current from left to right 
\mbox{$I_M=\lim\limits_{t\to\infty} \frac{d}{dt} \expval{N_B^{(R)}}$}, its zero-frequency noise 
\mbox{$S_M=\lim\limits_{t\to\infty} \frac{d}{dt} \left[\expval{\left(N_B^{(R)}\right)^2}-\left(\expval{N_B^{(R)}}\right)^2\right]$},
 and also the stationary energy current from left to right \mbox{$I_E=\lim\limits_{t\to\infty} \frac{d}{dt} \expval{H_B^{(R)}}$} can all be expressed by one-dimensional integrals
\begin{align}\label{EQ:curnoisecur}
I_M &= \int \frac{d\omega}{2\pi} T(\omega) \left[f_L(\omega) - f_R(\omega)\right]\,,\nn
S_M &= \int \frac{d\omega}{2\pi} \Big\{T(\omega) \left[\sum_\nu f_\nu(\omega)(1-f_\nu(\omega))\right]\nn
&\qquad+ T(\omega)[1-T(\omega)]\left[f_L(\omega) - f_R(\omega)\right]^2\Big\}\,,\nn
I_E &= \int \frac{d\omega}{2\pi} \omega T(\omega) \left[f_L(\omega) - f_R(\omega)\right]\,,
\end{align}
where the transmission function $0 \le T(\omega) \le 1$ describes how energy and particles are transmitted through the system at energy $\omega$ and $f_\nu(\omega) = [e^{\beta_\nu(\omega-\mu_\nu)}+1]^{-1}$ denotes the Fermi function of lead $\nu$ held at inverse temperature $\beta_\nu$ and chemical potential $\mu_\nu$.
We stress that these expressions remain valid at arbitrary system-reservoir coupling strength.
The noise splits into two contributions~\cite{blanter2000a}: The thermal noise (first line in $S_M$), which vanishes at zero temperature due to the $f_\nu(\omega)[1-f_\nu(\omega)]$-dependence of the integrand, and the shot or non-equilibrium noise (second line in $S_M$), that is suppressed in near-equilibrium scenarios where $f_L(\omega) \approx f_R(\omega)$.
We also note that the shot noise can be suppressed when the transmission $T(\omega)$ is either zero or close to unity throughout the region where $f_L(\omega)\neq f_R(\omega)$ (also called transport window).

From the Landauer equations above, we can already conclude a few standard thermodynamic implications. 
When we construct the entropy production rate $\sigma$ from the energy currents $I_E^{(\nu)}$ and matter currents $I_M^{(\nu)}$ leaving reservoir $\nu$, we find by using 
conservation of stationary currents $I_E = I_E^{(L)} = -I_E^{(R)}$ and $I_M = I_M^{(L)}=-I_M^{(R)}$ that
\begin{align}\label{EQ:entprod}
\sigma &= -\beta_L \left(I_E^{(L)}-\mu_L I_M^{(L)}\right)-\beta_R \left(I_E^{(R)}-\mu_R I_M^{(R)}\right)\nn
&= (\beta_R-\beta_L) I_E + (\beta_L \mu_L - \beta_R \mu_R) I_M \ge 0\,,
\end{align}
where the inequality
can be seen~\cite{nenciu2007a,topp2015a} by the strict positivity of the resulting integrand when we insert~\eqref{EQ:curnoisecur}.
One can check that this inequality implies the usual common wisdoms: At equal temperatures $\beta_L=\beta_R=\beta$, the matter current will flow from the lead with higher chemical potential towards the lead with lower one, and at equal potentials $\mu_L=\mu_R=\mu$, the heat current $I_Q = I_E - \mu I_M$ will flow from hot to cold reservoir as in Clausius formulation of the second law, i.e., under reversal of the corresponding thermodynamic biases, the currents must change sign. 
A more interesting scenario arises however when both chemical potential and temperatures are different.
In this case, one may e.g. use the temperature gradient to drive a particle current against the potential bias (which generates chemical work by charging the reservoirs like a battery) or -- alternatively -- use the potential bias to pump heat from cold to hot (cooling the colder reservoir).
For these cases, one may use the above inequality to show that the efficiency of work extraction from a thermal gradient and coefficient of performance for cooling via investing chemical work are both bound by their corresponding Carnot limits.
From more general considerations~\cite{esposito2010b} we would also expect that such an inequality must hold for all two-terminal models supporting stationary states for the central system and conserved stationary energy and particle currents in each terminal (i.e., even if interactions were included~\cite{meir1992a}).

Experimentally, the transmission function $T(\omega)$ is accessible via the current as is formally exemplified for $\beta_L=\beta_R=\beta$ and $\mu_L=\bar\mu+V/2$ and $\mu_R=\bar\mu-V/2$ by the relation for the conductance at low temperatures and small bias voltage
\begin{align}
\lim_{\beta\to\infty}  \lim_{V\to 0} \frac{dI_M}{dV} = \frac{T(\bar\mu)}{2\pi}\,,
\end{align}
which can be understood formally from $\lim_{\beta\to\infty}  \lim_{V\to 0} [f_L(\omega)-f_R(\omega)] = \delta(\omega-\bar\mu)$.
That is, at low temperature, peaks of the transmission function map to steps in the current-voltage characteristics. 

Numerically, the transmission function can be obtained by computing a trace over the lesser Greens function and the coupling matrices to the reservoir. 
Particularly, using the specifics of our model where we couple between system and reservoir only at a single site, this expression collapses to the computation of extremal matrix elements of the central (by ''central'' we understand all sites for which the Hamiltonian differs from a homogeneous chain -- the green circles in Fig.~\ref{FIG:model_sketch}) retarded Greens function
\begin{align}
T(\omega) = \abs{G_{0,N+1}(\omega)}^2 (4\tau_0^2-\omega^2)\,.
\end{align}
Above, we have also used that the remaining coupling strength to the residual reservoirs (white dots in Fig.~\ref{FIG:model_sketch}) 
are just $\Gamma_L(\omega)=\Gamma_R(\omega)=\sqrt{4\tau_0^2-\omega^2} \Theta(4\tau_0^2-\omega^2)$, as the $\tau_\nu$ are contained inside the central system.
The computation of the central Greens function is outlined in Sec.~\ref{SEC:greensfunction} below.

\subsection{Stationary system quantities}

The other matrix elements of the central Greens function also determine other expectation values, for example we have
for the two sites described by double-indices $i,j\in\{1,\ldots,N\}$ with $i=(i_x-1) N_y+i_y$ and $j=(j_x-1)N_y+j_y$ the expression
\begin{align}\label{EQ:occupation}
\expval{c_i^\dagger c_j}_{\rm SS} &= \int \frac{d\omega}{2\pi} \Big[
G_{j,1}(\omega) G_{i,1}^*(\omega) \Gamma_L(\omega) f_L(\omega)\nn
&\qquad + G_{j,N}(\omega) G_{i,N}^*(\omega) \Gamma_R(\omega) f_R(\omega)\Big]\,,
\end{align}
such that the local site occupation of site $i=j$ can be obtained directly.
However, for $i\neq j$ this also allows to compute the net particle current flowing along the bond connecting sites $i$ and $j$.
For a generic tunneling Hamiltonian of the form $H_S = \sum_{ij} (t_{ij} c_i^\dagger c_j + t_{ij}^* c_j^\dagger c_i)$ -- compare Eq.~\eqref{EQ:ham_hofstadter} -- one 
eventually obtains 
\begin{align}\label{EQ:bondcurrent}
I_{i\to j} = 2 \Im \left\{t_{ij} \expval{c_i^\dagger c_j}_{\rm SS}\right\}\,.
\end{align}
%
%
We checked numerically that the occupations obey $0 \le n_i \le 1$ and furthermore that the currents are conserved (e.g. that $I_M=I_{(1,1)\to (1,2)}+I_{(1,1)\to (2,1)}$ and similar).

\subsection{Central Greens function}\label{SEC:greensfunction}

To evaluate the central retarded Greens function, we briefly expose the non-equilibrium Greens function formalism~\cite{economou2006,haug2008,wang2014a}.
The computation of the full (retarded) Greens function $G(\omega)=\lim_{\delta\to 0} (\omega+\ii\delta-H)^{-1}$ requires a known exact solution of a free Greens function $G_0(\omega)=\lim_{\delta\to 0} (\omega+\ii\delta-H_0)^{-1}$ that we take as an infinite chain with hopping amplitude $\tau_0$
\begin{align}
H_0 = \tau_0 \sum_{n=-\infty}^{+\infty} [c_n^\dagger c_{n+1} + c_{n+1}^\dagger c_n]\,,
\end{align}
which runs along the orange line in Fig.~\ref{FIG:model_sketch}.
For this problem, the matrix elements of the retarded Greens function can be obtained by direct diagonalization~\cite{economou2006,odashima2017a,boehling2018a}
\begin{align}\label{EQ:matel_freegf}
(G_0)_{\ell,m}(\omega) = \frac{-\ii}{\sqrt{4 \tau_0^2-\omega^2}}\left(\frac{\omega}{2\tau_0}-\ii \sqrt{1-\frac{\omega^2}{4\tau_0^2}}\right)^{\abs{\ell-m}}\,,
\end{align}
where we assumed $4 \tau_0^2 > \omega^2$ (otherwise all matrix elements vanish).
Now, splitting the total Hamiltonian as $H=H_0+H_1$, the full retarded Greens function can be computed via a Dyson series expansion
\begin{align}\label{EQ:dysonseries}
G(\omega) &= [\omega-H_0-H_1]^{-1}\nn
&= [\f{1}-G_0(\omega)H_1]^{-1} G_0(\omega)\,.
\end{align}
In particular, $H_1$ has a local structure only compared to the free Greens function $G_0(\omega)$
\begin{align}
H_1&= \left(\begin{array}{ccc}
\f{0} & \f{0} & \f{0}\\
\f{0} & \bar H_1 & \f{0}\\
\f{0} & \f{0} & \f{0}
\end{array}\right)\,,\nn
G_0(\omega) &= \left(\begin{array}{ccc}
G_0^{LL} & G_0^{LC} & G_0^{LR}\\
G_0^{CL} & G_0^{CC} & G_0^{CR}\\
G_0^{RL} & G_0^{RC} & G_0^{RR}
\end{array}\right)\,,
\end{align}
from which it follows that the inverse in~\eqref{EQ:dysonseries} can be expressed as
\begin{align}
[\f{1}-G_0(\omega)H_1]^{-1} &= \left(\begin{array}{ccc}
\f{1} & G_0^{LC} \bar H_1 [\f{1}-G_0^{CC} \bar H_1]^{-1} & \f{0}\\
\f{0} & [\f{1}-G_0^{CC} \bar H_1]^{-1} & \f{0}\\
\f{0} & G_0^{RC} \bar H_1 [\f{1}-G_0^{CC} \bar H_1]^{-1} & \f{1}
\end{array}\right)\,.
\end{align}
Eventually, we obtain that the full retarded Greens function is given by 
\begin{align}
G(\omega) &= \left(\begin{array}{c|c|c}
\hdots & \hdots & \hdots\\
\hline
\hdots & [\f{1}-G_0^{CC} \bar H_1]^{-1} G_0^{CC} & \ldots\\
\hline
\hdots & \hdots & \hdots
\end{array}\right)\,,
\end{align}
and in particular for the evaluation of its central part, only an $(N+2)\times(N+2)$-dimensional matrix has to be inverted (numerically, one may prefer to solve the corresponding linear system instead).

\section{Results}\label{SEC:results}

\subsection{Near-unit transmission}

While the transmission through tight-binding models of 2d topological insulators has been investigated before (see e.g.~\cite{fremling2020a,fischer2021a}), we would like to add on the aspect of the system-reservoir coupling strength and an actual demonstration of robustness with respect to perturbations here.

\begin{figure}[ht]
\begin{tabular}{ccc}
\includegraphics[width=0.23\textwidth]{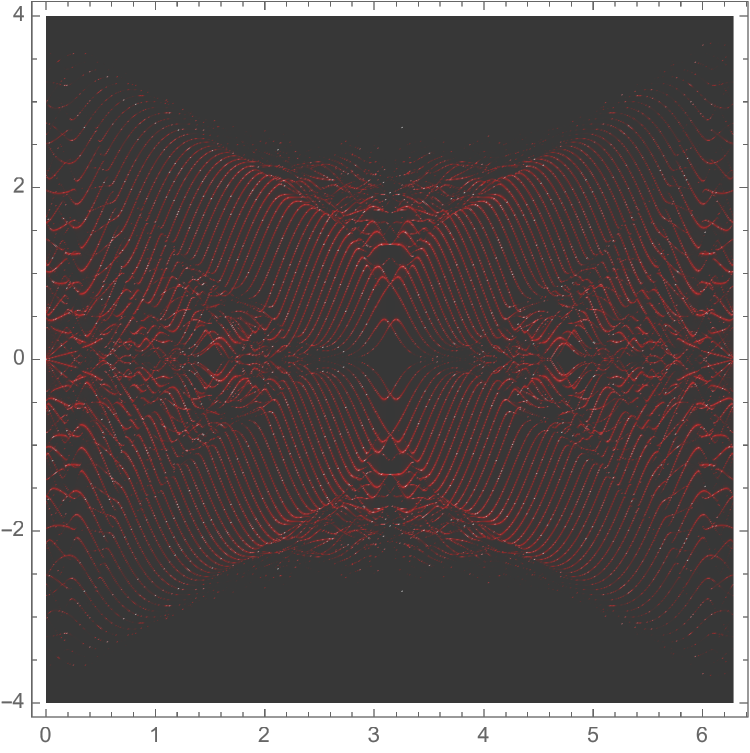} &
\includegraphics[width=0.23\textwidth]{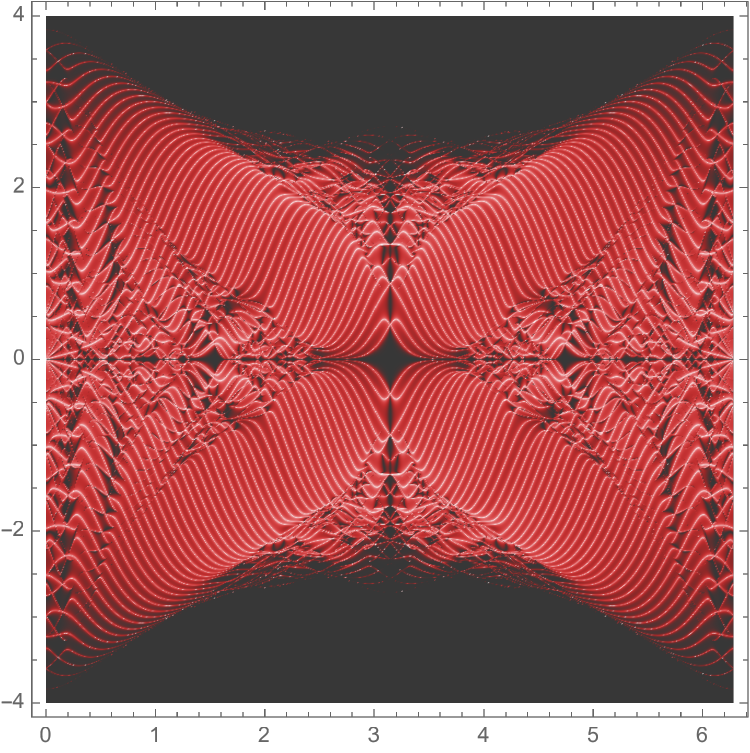} &
\includegraphics[height=0.22\textwidth]{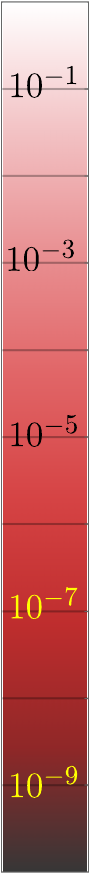}\\
\includegraphics[width=0.23\textwidth]{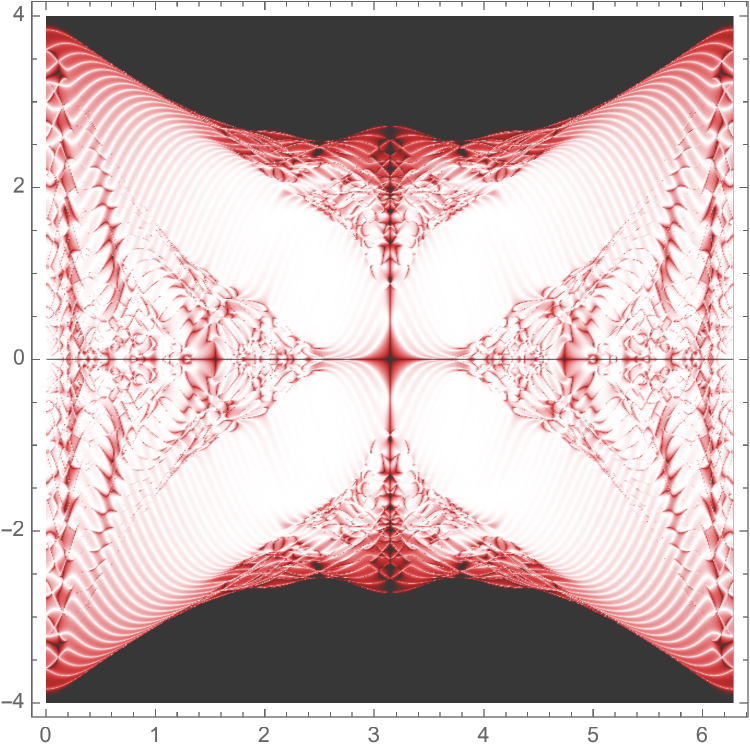} &
\includegraphics[width=0.23\textwidth]{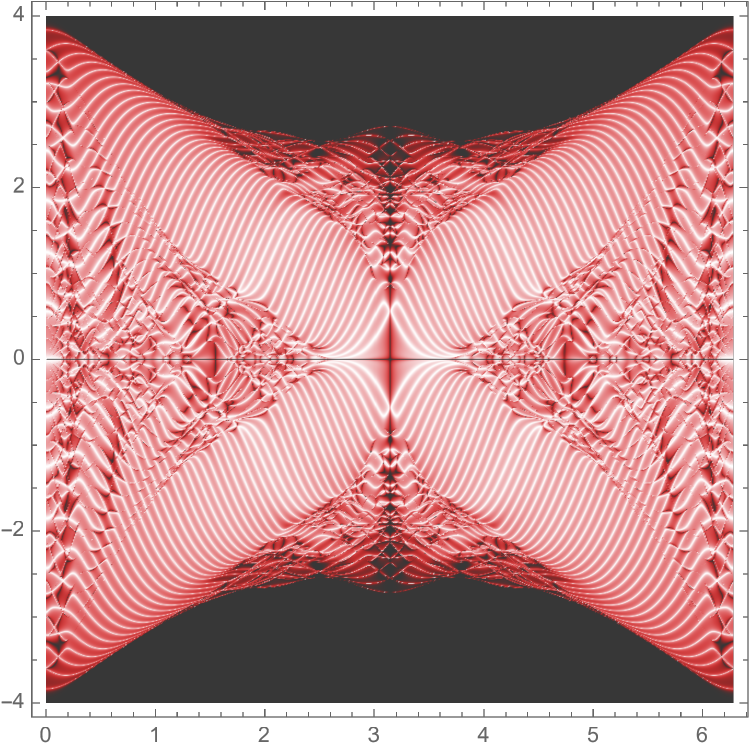}
\end{tabular}
\caption{\label{FIG:transmission}
Density plots of transmission (note the log-scale colorbar) versus magnetic field ${\cal B}=(e/\hbar) B \Delta^2$ (horizontal axis) and frequency $\omega/\tau$ (vertical axis). 
At weak coupling (top left), the transmission is in general very small and has extremely narrow (hardly visible) peaks that reproduce the Hofstadter butterfly spectrum from Fig.~\ref{FIG:butterfly}. Already at intermediate coupling (top right), the transmission is much larger. At specific coupling strengths, the transmission of the boundary modes is extremely close to one (bottom left), whereas at even larger coupling, the transmission reduces again (bottom right). 
Parameters: $N_x=N_y=10$, $\tau_0=10\tau$, $\tau_L=\tau_R=\tau_c$, where from top left to bottom right $\tau_c =(0.1, 0.5, 4.0, 10.0) \tau$.
}
\end{figure}

Numerically, we find that the transmission function for our system is strongly dependent on the system-reservoir coupling strength $\tau_\nu$, see Fig.~\ref{FIG:transmission}, where we display density plots of $T(\omega)$ vs. magnetic field strength ${\cal B}$ (horizontal axis) and frequency $\omega$ (vertical axis).
At weak coupling strengths (top left), ultra-narrow peaks in the transition just resemble the system energy eigenstates, compare with the orange curves in Fig.~\ref{FIG:butterfly}.
When coupling strengths are increased (top right), these peaks will broaden, until one finds whole regions of near-unit transmission (bottom left) at an optimal coupling strength of 
$\tau_L=\tau_R=4 \tau$ in the butterfly wings.
At even larger coupling strengths, the transmission decreases again but still exhibits the butterfly structure (bottom right).

\begin{figure}[ht]
\includegraphics[width=0.45\textwidth]{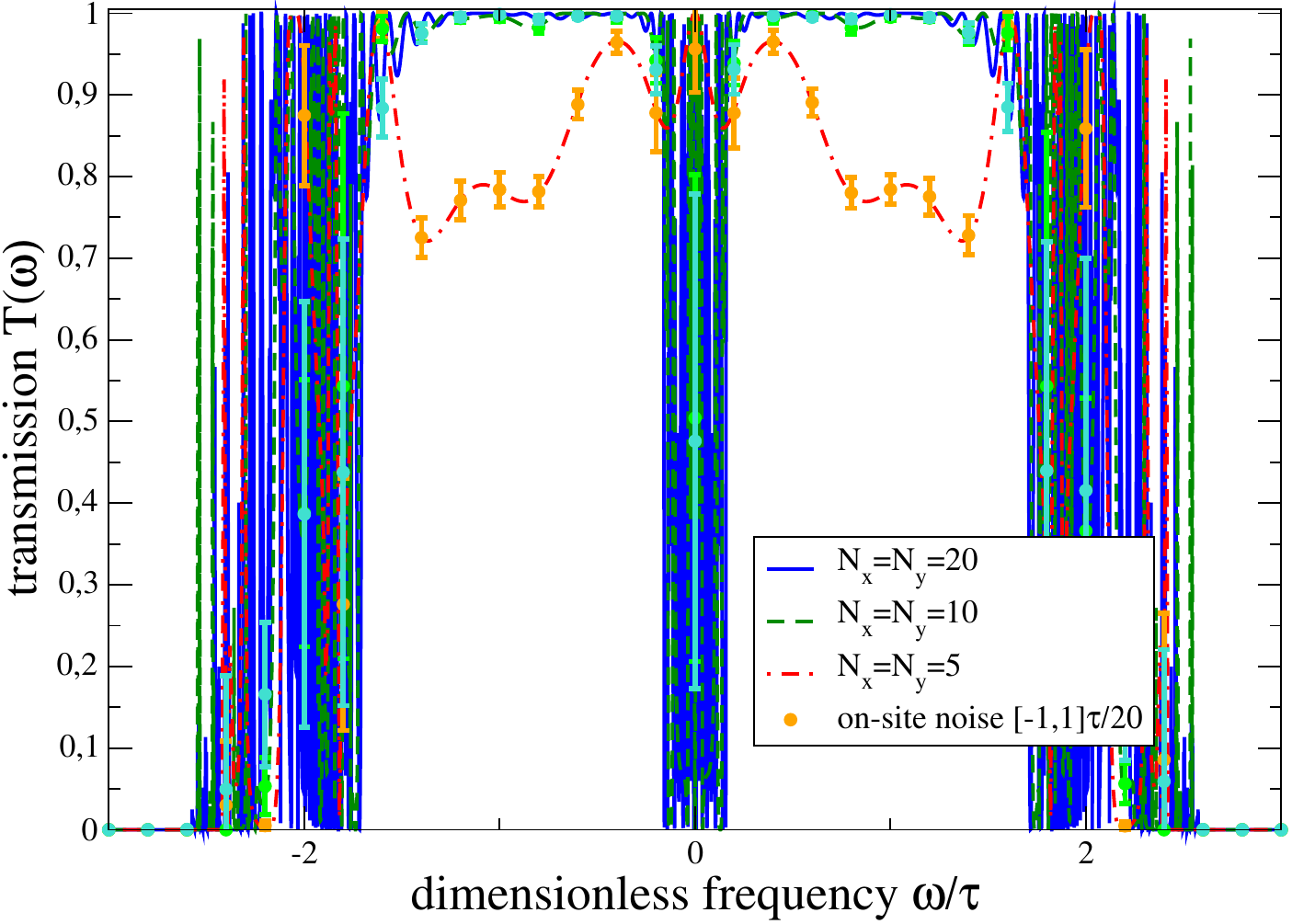}
\caption{\label{FIG:transmissionsweetspot}
Plot of the transmission $T(\omega)$ vs. frequency $\omega/\tau$ for ${\cal B}=2.5$ (passing along the arrow symbols in Fig.~\ref{FIG:butterfly}). 
With increasing system size (red dash-dotted to green dashed to blue solid), the transmission exhibits plateaus where the transmission is near unity.
Here, transport is mainly carried by topologically protected boundary modes. 
Outside these plateaus, the transmission shows for larger system sizes a lot of very narrow spikes that result from bulk state transmission.
When we add moderate on-site noise of the form~\eqref{EQ:onsitenoise} with $\alpha=0.1$, we find that the large-transmission plateaus are hardly affected (small error bars), whereas smaller transmissions suffer rather large variations (circle symbols of lighter colors result from averages over 100 noise realizations with error bars denoting one standard deviation).
Other parameters: $\tau_0 = 10 \tau$, $\tau_L=\tau_R=4\tau$.
}
\end{figure}

In the following, we want to explore some uses of large transmission function and will therefore concentrate on the regime $\tau_L=\tau_R=4\tau$ and ${\cal B} = 2.5$.
For these parameters, we plot the transmission for different system sizes $N=N_x N_y$ in Fig.~\ref{FIG:transmissionsweetspot}.
The cut is along the vertical line covering the red and green arrows in Fig.~\ref{FIG:butterfly} and shows that in the butterfly wings, plateaus of unit transmission become more pronounced the larger the system is.
Away from these plateaus, the transmission functions for large systems exhibit many narrow spikes that correspond to eigenstates concentrated in the bulk.

When we add fluctuations to the on-site energies to the system Hamiltonian $H_S \to H_S + \Delta H_S$ with
\begin{align}\label{EQ:onsitenoise}
\Delta H_S = \sum_{a=1}^{N_x} \sum_{b=1}^{N_y} \epsilon_{a,b} c_{a,b}^\dagger c_{a,b}\,,
\end{align}
where each of the on-site energies is independently randomly chosen from a uniform distribution in the interval $\epsilon_{a,b} \in [-\tau/2, +\tau/2]\alpha$,
the transmission function behaves quite different depending on whether it is evaluated for boundary modes (near unit transmission) or bulk modes (regions with a lot of spikes).
Boundary modes are hardly affected by the onsite-noise, and the corresponding statistical variance (error bars) is small (even smaller so for larger system sizes).
This is different for the bulk modes, which have very narrow spikes that shift a bit in position. 
Therefore, averaging over them yields a much smaller average transmission and larger variance in these regimes.
This behaviour is expected because the boundary modes are topologically protected: The random onsite energies~\eqref{EQ:onsitenoise} break the chiral symmetry, but they still respect the gauge symmetry that is required for the protection of the boundary modes.

We note that analogous near-unit transmission plateaus can also be found for Aharonov-Bohm rings~\cite{haack2019a} -- which boundary-mode transport effectively implements -- and that the corresponding applications that we investigate below can in principle also be found there.

\subsection{Bond currents}

\begin{figure}[ht]
\begin{tabular}{cc}
\includegraphics[width=0.22\textwidth]{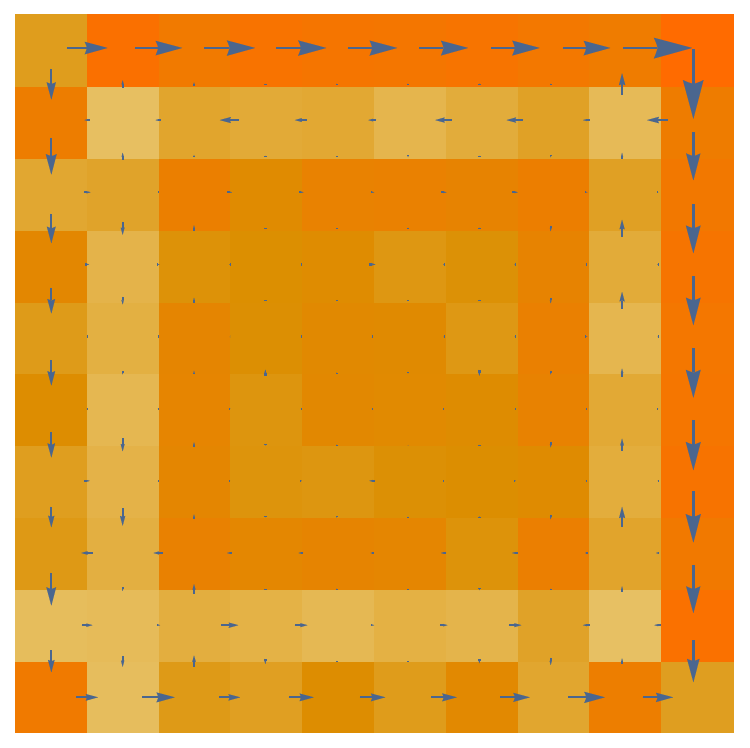} & \includegraphics[width=0.22\textwidth]{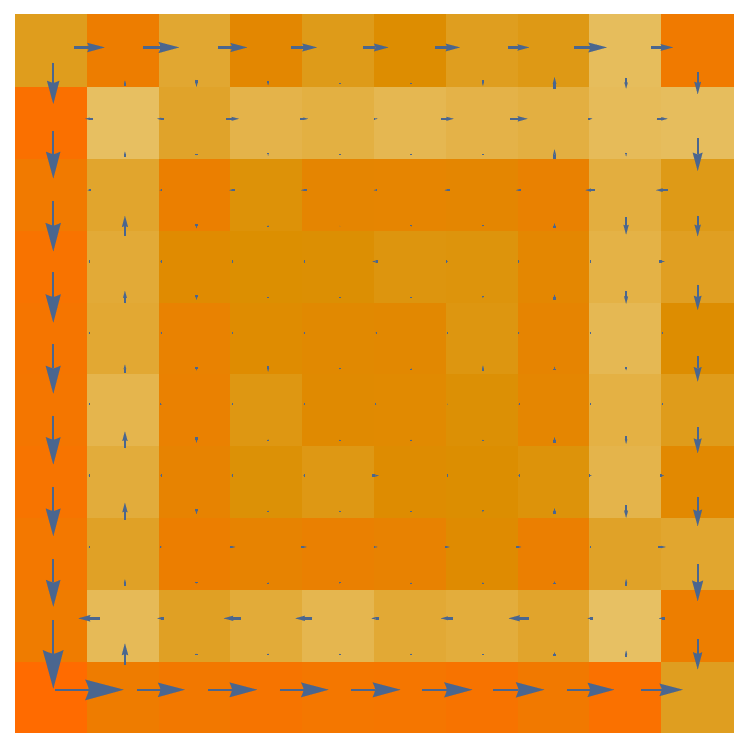}\\
\includegraphics[width=0.22\textwidth]{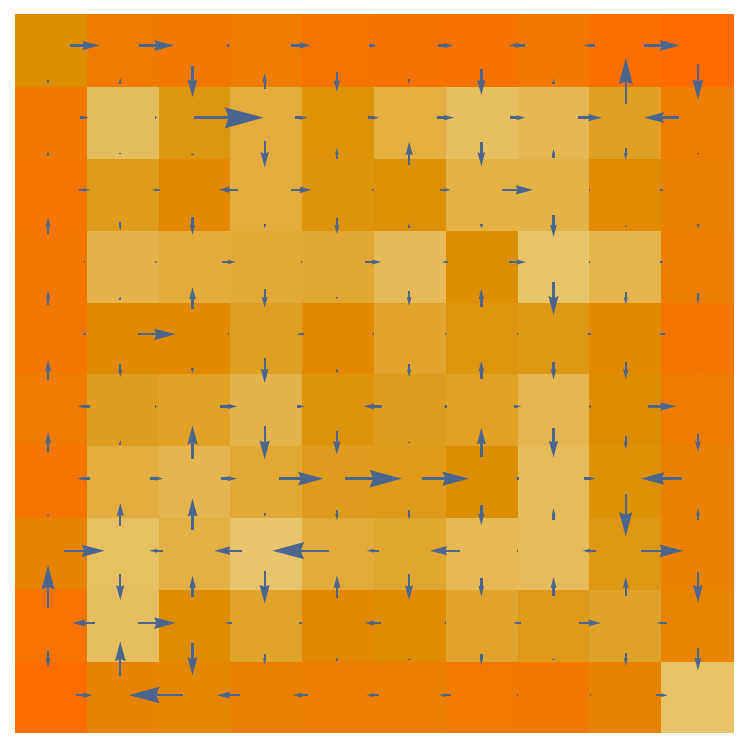} & \includegraphics[width=0.22\textwidth]{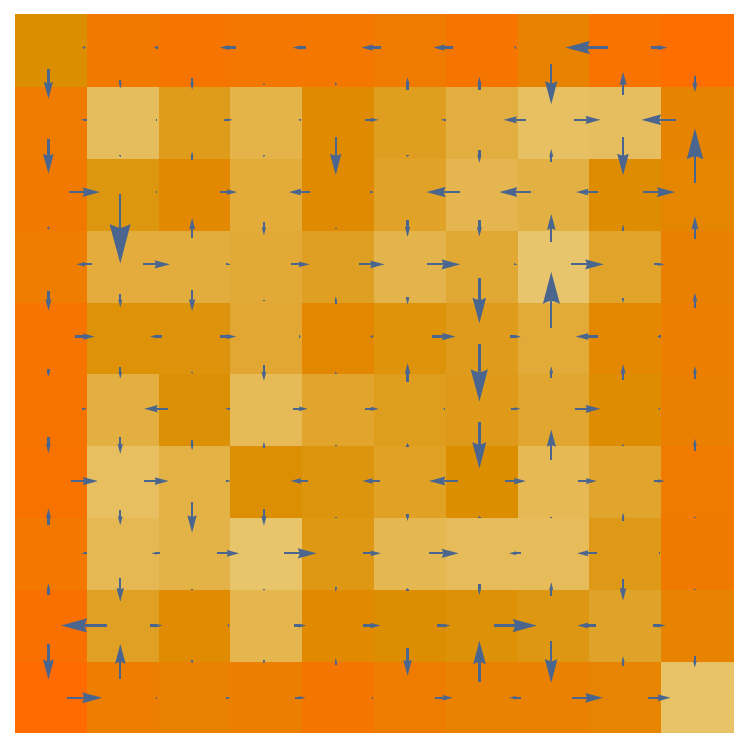}
\end{tabular}
\caption{\label{FIG:occupations}
Plots of stationary occupations (background colors) and bond currents (arrow lengths correlates with strength) for different average potentials and magnetic fields (top vs. bottom row) under reversal of the respective magnetic field ${\cal B}$ (left vs right column).
The top row shows the occupation for average potential $\bar\mu=(\mu_L+\mu_R)/2=+1.0\tau$ for ${\cal B}=\pm 2.5$ (green rectangles in Fig.~\ref{FIG:butterfly}) and thus exhibits the transport mediated by topological boundary modes (total current $I_M \approx 0.079 \tau$).
The bottom row is for average potential $\bar\mu=2$ and similarly contrasts ${\cal B}=\pm 3$ (yellow rectangles in Fig.~\ref{FIG:butterfly}) the transport through bulk modes as well as ring currents without net charge transfer (total current $I_M \approx 0.024 \tau$).
Other parameters: $\tau_0=10\tau$, $\tau_L=\tau_R=4\tau$, $\tau\beta_L=\tau\beta_R=10$, and $V=\mu_L-\mu_R=0.5\tau$.
}
\end{figure}

At a finite bias and equal temperatures, we plot the stationary occupations from~\eqref{EQ:occupation} and bond currents from~\eqref{EQ:bondcurrent} for different average chemical potentials and magnetic fields in Fig.~\ref{FIG:occupations}.
Increased stationary occupations indicate the paths the currents are taking and coincide in case of boundary transport with the actual bond currents (top row in Fig.~\ref{FIG:occupations} corresponds to the green rectangles in Fig.~\ref{FIG:butterfly}), thus supporting the view that with appropriate potential choices one may selectively excite surface modes.
Additionally, one can see in the top row that the two arms of the current are differently strong, depending on the direction of the magnetic field. 
When we choose parameters that dominantly excite bulk modes (the bottom row in Fig.~\ref{FIG:occupations} corresponds to the yellow rectangles in Fig.~\ref{FIG:butterfly}), the currents along the boundary contribute little to net transport, whereas currents through the bulk are much larger.

\subsection{Precise and robust transport statistics}

From the counting statistics~\eqref{EQ:curnoisecur}, we could estimate the long-term number of transferred particles after large times $t$ as $\expval{n}_t \approx I_M t$ and
also their variance as $\expval{n^2}_t-\expval{n}_t^2 \approx S_M t$.
When the chemical potentials are tuned such that the transport window (the frequencies for which $f_L(\omega) \neq f_R(\omega)$) lies inside the large-transmission plateaus where $T(\omega)\approx 1$, the shot-noise contribution to the noise~\eqref{EQ:curnoisecur} will be strongly suppressed.
Likewise, the thermal noise can also be suppressed by selecting appropriately small temperatures, and by choosing both potentials and temperatures one may implement low-noise transport ($S_M \gtrsim 0$ and $I_M > 0$ such that the Fano factor $F=S_M/I_M$ is suppressed).
Given a precise clock, a useful application of this is the definition of hyperaccurate current standards~\cite{timpanaro2023a,timpanaro2025a,khandelwal2025a,sobrino2026a}.

One may also turn the question around and ask for the average time $\expval{t}_n$ at which a certain given particle number $n$ is reached first, which falls in the class of first-passage-time problems~\cite{vankampen1981}, see e.g. Ref.~\cite{singh2019a} for corresponding electronic transport experiments.
Given a reliable mechanism to count the number of passed electrons $n$ e.g. via capacitive or optical means~\cite{flindt2009a,kurzmann2019a}, this naturally implements a clock~\cite{erker2017a}.
In general, the computation of the passage times from the counting statistics is non-trivial already in the Markovian case~\cite{menczel2026a} or for unidirectional transport~\cite{albert2011a} but under Markovianity assumptions the cumulants of the passage time distribution can be estimated from the cumulants of the counting statistics (see e.g. Eqns.~(34) and~(35) in Ref.~\cite{ptaszynski2018b}) as 
\begin{align}\label{EQ:cumconv}
\expval{t}_n &\approx \frac{n}{I_M}\,,\nn
\expval{t^2}_n-\expval{t}_n^2 &\approx \frac{S_M}{I_M^2} \expval{t}_n\,.
\end{align}

Consequently, the thermodynamic uncertainty relation (TUR)~\cite{barato2015a} establishing a bound on noise, current, and entropy production rate for dissipative Markovian processes
\begin{align}\label{EQ:tur}
\frac{S_M}{I_M^2} \sigma \ge 2\,,
\end{align}
or looser bounds established for dissipative coherent Markovian processes for small thermal and potential gradients~\cite{guarnieri2019a} (LTUR)
\begin{align}\label{EQ:ltur}
\frac{S_M}{I_M^2} \sigma \ge 1\,,
\end{align}
would both have consequences for the precision of clocks: 
Inserting relations~\eqref{EQ:cumconv} in the precision we obtain 
\begin{align}
p=\frac{\expval{t}_n^2}{\expval{t^2}_n - \expval{t}_n^2} \approx \frac{n}{I_M} \frac{I_M^2}{S_M}\,,
\end{align}
and when the second factor is upper-bounded by the entropy production rate (LTUR) or half the entropy production rate (TUR), we find that in both cases precision is upper-bounded by the total entropy production.
In contrast, when the bounds~\eqref{EQ:tur} and~\eqref{EQ:ltur} are not respected (which could happen e.g. in intrinsically non-Markovian regimes and/or very far from equilibrium), high precision can be reached with less entropy production, which is compatible with the observation that thermodynamics does not limit precision~\cite{meier2025a}.

\begin{figure}[ht]
\includegraphics[width=0.45\textwidth]{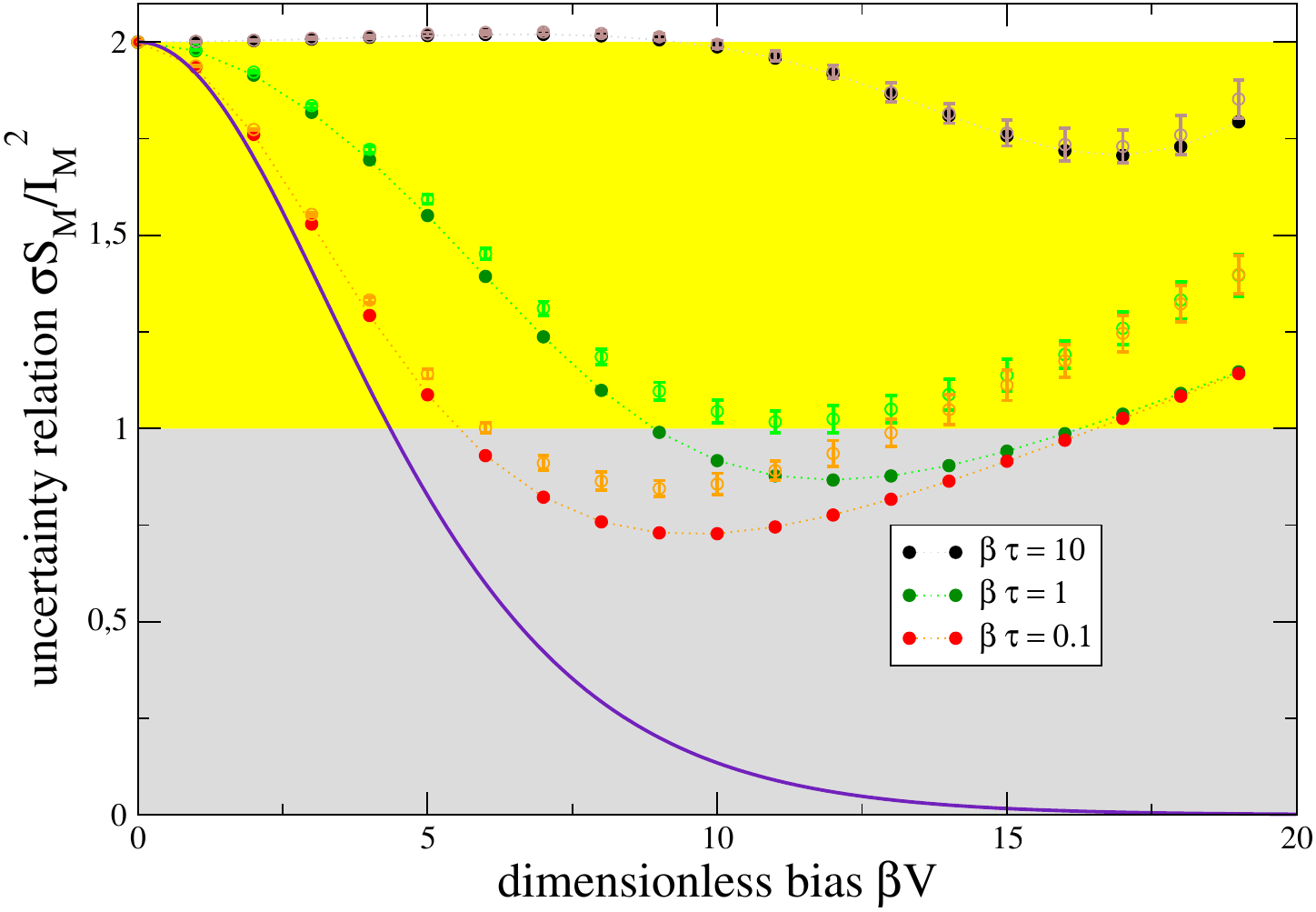}
\caption{\label{FIG:tur}Plot of the uncertainty relation vs. dimensionless bias voltage. 
While at equilibrium ($V=0$), the TUR~\eqref{EQ:tur} is respected, it can be easily broken when a bias voltage is applied (yellow background).
This even applies to the LTUR bound~\eqref{EQ:ltur} that is indicated by a gray background.
The CTUR bound~\eqref{EQ:ctur} is depicted by the bold violet curve.
Lighter symbols with error bars have been calculated from the statistical average over 100 realizations of onsite-noise~\eqref{EQ:onsitenoise} with $\alpha=0.1$.
Parameters $N_x=N_y=10$, $\bar\mu=1.0\tau$, $\tau_0=10\tau$, ${\cal B}=2.5$, $\tau_L=\tau_R=4\tau$ (compare green rectangle with arrows in Fig.~\ref{FIG:butterfly}).
}
\end{figure}

By appropriately tuning the bias window for our model, we find a clear violation of the TUR and LTUR, see Fig.~\ref{FIG:tur}.
To do so, we choose $\mu_L = \bar\mu+V/2$ and $\mu_R=\bar\mu-V/2$ with $\bar\mu = \tau$ such that the center of the transport window is approximately in the center
of the largest butterfly wing (green rectangle in Fig.~\ref{FIG:butterfly}).
At low temperatures, we indeed find a violation of the TUR bound away from the equilibrium regime (black symbols).
When we consider higher temperatures (green and red symbols), larger bias voltages are required but even stronger violations of TUR and LTUR bounds are observed.
Our interpretation of this is that in the latter case, the whole transmission interval of the system contributes instead of just one butterfly wing.
Still, it requires also in this case that the transmission function reaches near unity.
Additionally, we remark that the breaking of this bound does not require precise fine-tuning, as adding onsite-noise of the form~\eqref{EQ:onsitenoise} to the model does not substantially change the result (lighter symbols with error bars). 
This differs from previous results on quantum dot chains that required precise fine-tuning to beat TURs~\cite{agarwalla2018a,ehrlich2021a,brandner2025a} and can possibly be related to multi-particle processes relevant for stronger couplings~\cite{ohnmacht2026a}.

Finally, we may also comment on the coherent TUR (CTUR) bound recently derived (albeit under the assumptions of absent magnetic fields)~\cite{brandner2025a,brandner2025b}
\begin{align}\label{EQ:cturfull}
\frac{S_M}{I_M} \sinh\left(\frac{\sigma}{2 I_M}\right) \ge 1\,.
\end{align}
For the case of equal temperatures, where $\sigma=\beta V I_M$, this can be rearranged as
\begin{align}\label{EQ:ctur}
\frac{S_M}{I_M^2} \sigma \ge \frac{\beta V}{\sinh\left(\frac{\beta V}{2}\right)}\,,
\end{align}
the r.h.s. of which we show as the bold violet curve in Fig.~\ref{FIG:tur}.
An optimally placed and optimally sized boxcar transmission would actually saturate this bound.
One can see that this bound is quite tight for small voltages, indicating that our setup without any particular fine-tuning in the system already implements a quite optimal boxcar-type transmission function.

\subsection{Performant and robust thermoelectrics}

As it has been pointed out that boxcar-shaped transmission functions, which assume value $1$ in a certain interval and vanish elsewhere, may be very beneficial for thermoelectrics~\cite{whitney2014a}, we analyze the suitability of our model for these applications.
Without loss of generality, we will in this section consider the case $\beta_L > \beta_R$ (cold left and hot right reservoir) and $\mu_L>\mu_R$ (positive bias $V=\mu_L-\mu_R>0$).
Let us first find parameters for which we can expect positive power and positive cooling current, respectively.

\begin{figure}[ht]
\includegraphics[width=0.5\textwidth]{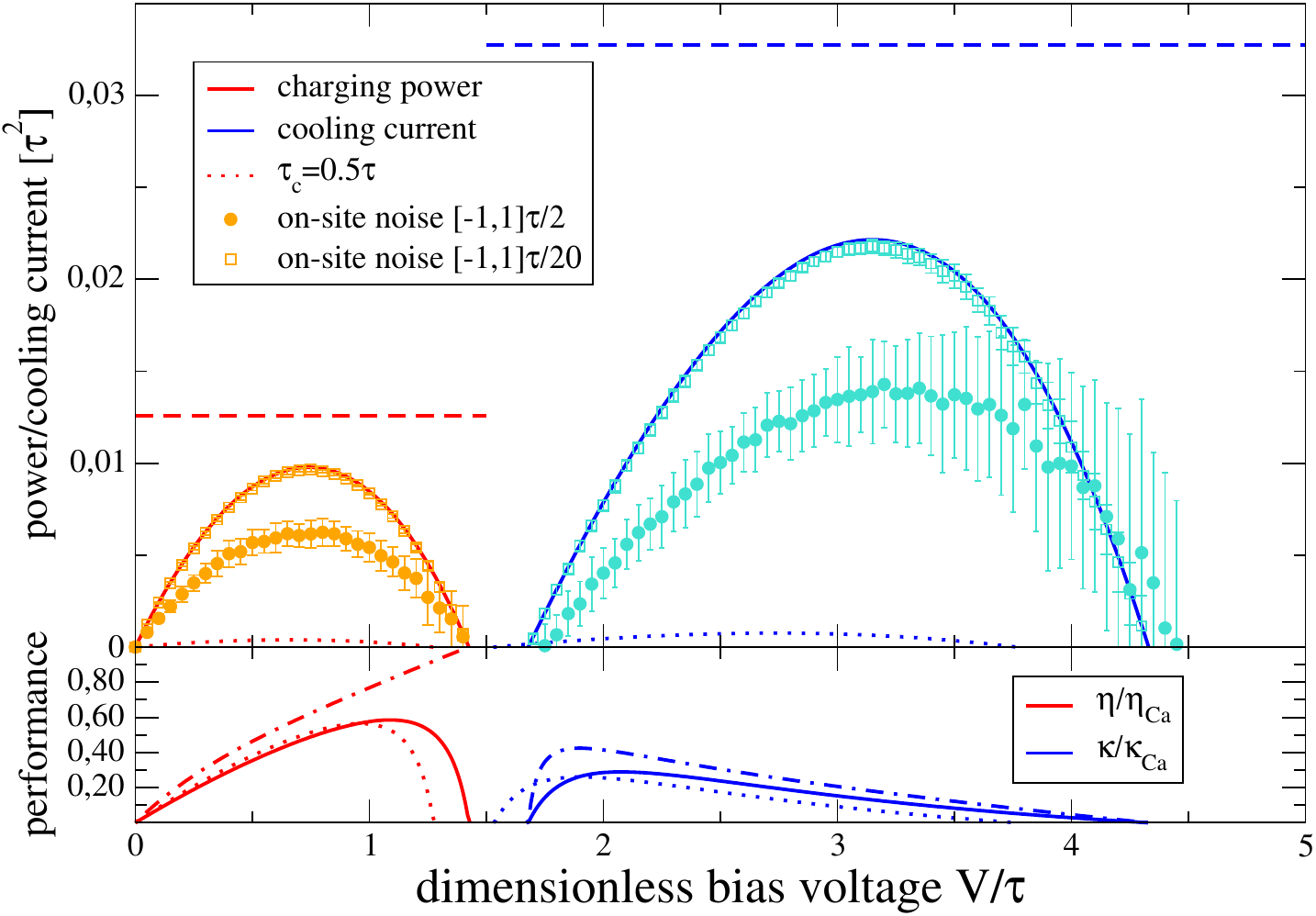}
\caption{\label{FIG:currentsandefficiency}
The top panel shows dimensionless power $P=-I_M V$ (solid red) and cooling current $I_Q = I_E - \mu_L I_M$ (solid blue) vs. dimensionless bias voltage. 
Each operational regime has an optimal bias voltage.
Choosing the optimal system-reservoir coupling strength greatly enhances transmission and thereby charging power and cooling current (solid vs. dotted curves).
The performance is also quite robust with respect to variations of the on-site energies~\eqref{EQ:onsitenoise}: Symbols and error bars denote average and statistical error of 100 realizations 
with $\alpha=0.1$ (hollow symbols with smaller error bars) and $\alpha=1.0$ (filled symbols with larger error bars).
Dashed lines indicate maximum power (red) and maximum cooling current (blue) that are attainable for boxcar transmission functions by varying $V$ and $\bar\mu$ at fixed temperatures.
The bottom panel demonstrates that the corresponding efficiency (red) and coefficient of performance (blue) well respect tighter-than Carnot bounds discussed in App.~\ref{APP:performancebounds} (dash-dotted curves).
Parameters: $\tau_0 = 10\tau$, $\tau_c=\tau_L=\tau_R=4\tau$ (solid and dashed curves) or $\tau_c=\tau_L=\tau_R=0.5\tau$ (dotted curves), ${\cal B}=2.5$, $N_x=N_y=10$, $\bar\mu=-3.5\tau$, $\beta_R\tau=1$, $\beta_L=2\beta_R$ (compare red rectangle with arrows in Fig.~\ref{FIG:butterfly}).
}
\end{figure}

First, to achieve positive charging power $P=-V I_M =-V \int \frac{d\omega}{2\pi} T(\omega) [f_L(\omega)-f_R(\omega)] > 0$ (the sign ensures that power is extracted when the current flows against the bias), we require that the transmission is large where $f_L(\omega)<f_R(\omega)$ and small elsewhere. 
Parametrizing $\mu_L=\bar\mu+V/2$ and $\mu_R=\bar\mu-V/2$, one can solve the equation $f_L(\omega)=f_R(\omega)$, which leads to the conclusion that $T(\omega)$ should be large
within the ideal window $\omega \in [\bar\mu + \frac{\beta_L+\beta_R}{\beta_L-\beta_R} \frac{V}{2},\infty)$.
Under the above assumptions on the temperatures and potentials this can be fulfilled by choosing $\bar\mu$ smaller than the smallest eigenvalue of the system (compare red rectangle in Fig.~\ref{FIG:butterfly}).
Clearly, we would then expect a positive charging power starting immediately at $V>0$, and this is what we see in the solid red curve in the main panel of Fig.~\ref{FIG:currentsandefficiency} for the chosen parameters (compare also the red rectangle in Fig.~\ref{FIG:butterfly}).
As we increase the bias voltage $V$, the lower bound of the ideal transmission window also rises, and for large enough bias voltages a significant fraction of the transmission window will be outside the ideal interval, thus contribute negatively to the total power, and we find that the charging power assumes a maximum at a specific voltage.
Where power is positive, the corresponding efficiency is found by dividing by the heat current entering from the hot (right) reservoir, which is bound by the 
corresponding Carnot limit $\eta = P/I_Q^{(R)} = (\mu_L-\mu_R) I_M/(I_E-\mu_R I_M) \le 1 - \beta_R/\beta_L$ (solid red in bottom panel).
In the appendix~\ref{APP:performancebounds} we show that Eq.~\eqref{EQ:cturfull} implies a tighter bound (dash-dotted red in bottom panel).

Second, to achieve a positive cooling current $I_Q^{(L)} = I_E -\mu_L I_M = \int \frac{d\omega}{2\pi} T(\omega) [\omega-\bar\mu-\frac{V}{2}][f_L(\omega)-f_R(\omega)]>0$, the transmission should be large for frequencies where $\omega > \bar\mu+\frac{V}{2}$ and $f_L(\omega)>f_R(\omega)$. 
Together with the previous results, this defines that for positive cooling current, the ideal transmission window obeys $\omega\in\left[\bar\mu+\frac{V}{2}, \bar\mu + \frac{\beta_L+\beta_R}{\beta_L-\beta_R} \frac{V}{2}\right]$.
This is exclusive with the previous condition for charging power, we would expect the cooling current to turn positive once the bias voltage is large enough to satisfy this condition and then decrease again beyond an optimal value, and this behaviour is also found in Fig.~\ref{FIG:currentsandefficiency} (blue curve in main panel).
When we divide the cooling current by the invested chemical work $-P>0$, we obtain the coefficient of performance for cooling (blue curve in bottom panel), which is also bound by the corresponding Carnot limit $\kappa = I_Q^{(L)}/(-P) = (I_E-\mu_L I_M)/[(\mu_L-\mu_R)I_M] \le \beta_R/(\beta_L-\beta_R)$.
In App.~\ref{APP:performancebounds}, we derive a tighter bound based on~\eqref{EQ:cturfull}, which is shown by the dash-dotted blue curve.
There is of course also the possibility to heat the hot reservoir, which we do not discuss here.

In addition, we see that although efficiency and coefficient of performance are similar, the magnitude of power and cooling current are much larger for the unit transmission regime compared to the weak-coupling limit -- this is actually one example where quantum heat engines can benefit from the strong-coupling regime~\cite{ivander2022a,latune2023a}.
One should also note that we did not separately optimize cooling and charging here, such that we fall a bit short on the maximum limits of 
$P \le \left(\frac{\beta_L-\beta_R}{\beta_L \beta_R}\right)^2 \frac{0.31635}{2\pi}$ for charging (dashed red line) and
$I_Q \le \frac{\pi}{24 \beta_L^2}$ for cooling (dashed blue line) that hold for a boxcar transmission that is optimized over $\bar\mu$ and $V$, 
compare Eqns.~(D8) and (D2) of Ref.~\cite{ehrlich2021a}, respectively.

Furthermore, we find that small on-site noise of the form~\eqref{EQ:onsitenoise} is hardly affecting this behaviour (hollow symbols) and even when the on-site noise is comparable to the system tunneling amplitudes, the operational modes are still accessible.
This is also expected as the near unit transmission that we exploit here is carried by topologically protected boundary modes.
As a further advantage we also note that in comparison to 1d toy models for topological insulators~\cite{boehling2018a}, the currents are by orders of magnitudes larger.
This can be understood as charges can in higher dimensions travel along the boundaries (compare also Fig.~\ref{FIG:occupations} top row) and need not tunnel through the system.

\section{Summary and Outlook}\label{SEC:summary}

Using non-equilibrium Greens function techniques, we have analyzed electronic transport characteristics through a Hofstadter model.
We found that for optimal system-reservoir coupling strengths at a single site, the boundary modes exhibit near unit transmission 
plateaus which are for sufficient system size quite robust with respect to local fluctuations of on-site energies (or background charges), whereas the 
peaked transmission generated by bulk states is not.
The inherent topological protection transfers to respective applications.
For example, the clear breaking of the thermodynamic uncertainty relation shows that such model systems have the potential to implement precise and robust current standards and quantum clocks based on electron counting.
Furthermore, when both thermal and voltage biases are applied, our setup may function as a thermoelectric generator (using a thermal gradient to charge a battery) or refrigerator (using charge flow to cool a cold reservoir). 
We thus find that optimal system-reservoir coupling strengths may strongly enhance the performance of these devices by enabling the required unit-transmission plateaus. 

We hope that our study triggers further investigations on role of optimal system-reservoir couplings and the robustness of quantum devices.

\begin{acknowledgments}
The authors gratefully aknowledge support by the Deutsche Forschungsgemeinschaft (DFG, German Research Foundation)
through the Collaborative Research Center SFB 1242 “Nonequilibrium dynamics of condensed matter in the time domain” (Project-ID 278162697).
\end{acknowledgments}

\bibliographystyle{unsrtnat}
\bibliography{/home/schall96/literatur/postdoc/postdoc}

\appendix

\section{Diagonalization for ${\cal B}=0$ and ${\cal B}=\pi$}\label{APP:diagonalization}

In this appendix, we use the relations
\begin{align}
\delta_{k,q} &= \frac{2}{N+1} \sum_{n=1}^N \sin\left[\frac{\pi k n}{N+1}\right]\times\nn
&\qquad\qquad\times \sin\left[\frac{\pi q n}{N+1}\right]\,,\nn
-\delta_{k+q,N+1} &= \frac{2}{N+1} \sum_{n=1}^N (-1)^n\sin\left[\frac{\pi k n}{N+1}\right] \times\nn
&\qquad\qquad\times \sin\left[\frac{\pi q n}{N+1}\right]\,,\nn
\cos\left[\frac{\pi k}{N+1}\right] \delta_{k,q} &= \frac{1}{N+1} \sum_{n=1}^{N-1}\times\nn
&\quad\times\Big\{\sin\left[\frac{\pi k n}{N+1}\right] \sin\left[\frac{\pi q (n+1)}{N+1}\right]\nn
&\qquad+\sin\left[\frac{\pi k (n+1)}{N+1}\right] \sin\left[\frac{\pi q n}{N+1}\right]\Big\}
\end{align}
that hold for integer $1\le k,q \le N$, to diagonalize the Hofstadter Hamiltonian~\eqref{EQ:ham_hofstadter} for the special cases ${\cal B}=0$ and ${\cal B}=\pi$.

We introduce the unitary transformation
\begin{align}
c_{a,b} &= \sqrt{\frac{2}{N_x+1}} \sqrt{\frac{2}{N_y+1}} \sum_{k=1}^{N_x} \sum_{q=1}^{N_y}\times\nn
&\qquad\qquad\times \sin\left(\frac{\pi a k}{N_x+1}\right) \sin\left(\frac{\pi b q}{N_y+1}\right) d_{kq}
\end{align}
to new fermionic operators $d_{kq}$.

In absence of a magnetic field, this implies that 
\begin{align}
H &= \tau_x \sum_{a=1}^{N_x-1} \sum_{b=1}^{N_y} (c_{a,b}^\dagger c_{a+1,b} + c_{a+1,b}^\dagger c_{a,b})\nn
&\qquad+ \tau_y \sum_{a=1}^{N_x} \sum_{b=1}^{N_y-1} (c_{a,b}^\dagger c_{a,b+1} + c_{a,b+1}^\dagger c_{a,b})\nn
&= \sum_{k=1}^{N_x} \sum_{q=1}^{N_y} \Big[2 \tau_x\cos\left(\frac{\pi k}{N_x+1}\right)\nn
&\qquad\qquad + 2 \tau_y \cos\left(\frac{\pi q}{N_y+1}\right)\Big] d_{kq}^\dagger d_{kq}\,,
\end{align}
which in the continuum limit generates two overlapping bands in the interval $(-2\abs{\tau_x}-2\abs{\tau_y},+2\abs{\tau_x}+2\abs{\tau_y})$
-- consistent with the left boundary of Fig.~\ref{FIG:butterfly} for $\tau_x=\tau_y=\tau$.

The very same transformation does not yet fully diagonalize the case ${\cal B}=\pi$
\begin{align}
H &= \tau_x \sum_{a=1}^{N_x-1} \sum_{b=1}^{N_y} (c_{a,b}^\dagger c_{a+1,b} + c_{a+1,b}^\dagger c_{a,b})\nn
&\qquad+ \tau_y \sum_{a=1}^{N_x} \sum_{b=1}^{N_y-1} (-1)^a (c_{a,b}^\dagger c_{a,b+1} + c_{a,b+1}^\dagger c_{a,b})\nn
&= \sum_{k=1}^{N_x} \sum_{q=1}^{N_y} 2 \tau_x\cos\left(\frac{\pi k}{N_x+1}\right) d_{kq}^\dagger d_{kq}\nn
&\qquad+ \sum_{k=1}^{N_x} \sum_{q=1}^{N_y} 2 \tau_y \cos\left(\frac{\pi q}{N_y+1}\right) d_{kq}^\dagger d_{N_x+1-k,q}\nn
&= 2 \tau_x \sum_{k=1}^{\lfloor N_x/2 \rfloor} \sum_{q=1}^{N_y} \cos\left(\frac{\pi k}{N_x+1}\right)\times\nn
&\qquad\qquad\times(d_{kq}^\dagger d_{kq} - d_{N_x+1-k,q}^\dagger d_{N_x+1-k,q})\nn
&\qquad+2 \tau_y \sum_{k=1}^{\lfloor N_x/2 \rfloor} \sum_{q=1}^{N_y} \cos\left(\frac{\pi q}{N_y+1}\right)\times\nn
&\qquad\qquad\times(d_{kq}^\dagger d_{N_x+1-k,q} - d_{N_x+1-k,q}^\dagger d_{kq})\nn
&\qquad+\tau_y [1-(-1)^{N_x}] \sum_{q=1}^{N_y} \cos\left(\frac{\pi q}{N_y+1}\right)\times\nn
&\qquad\qquad\times d_{\frac{N_x+1}{2},q}^\dagger d_{\frac{N_x+1}{2},q}\,,
\end{align}
where the last line is only existent for odd $N_x$ (for which it is already in diagonal form).
However, we can perform an additional rotation between the modes $(k,q)$ and $(N_x+1-k,q)$, which eventually yields the eigenvalues (for the first two lines)
\begin{align}
\epsilon_{kq}^\pm &= \pm \sqrt{2} \Big\{\tau_x^2\left[1+ \cos\left[\frac{2\pi k}{N_x+1}\right]	\right]\nn
&\qquad+ \tau_y^2 \left[1+ \cos\left[\frac{2\pi q}{N_y+1}\right]\right]\Big\}^{1/2}\,,
\end{align}
such that in the continuum limit we have two bands ranging from $(-2\sqrt{\tau_x^2+\tau_y^2},0)$ and $(0,+2\sqrt{\tau_x^2+\tau_y^2})$, respectively.
This is consistent with what we see in Fig.~\ref{FIG:butterfly} for ${\cal B}=\pi$.
Furthermore, for odd $N_x$ these bands would overlap with 
a third band in the interval $(-2\tau_y,+2\tau_y)$.

\section{Tighter bounds on performance}\label{APP:performancebounds}

As the traditional TUR inequality bounds heat engine efficiency~\cite{pietzonka2018a} and also coefficient of performance~\cite{kloc2021a}, one may conjecture
that its violation in coherent thermoelectrics enables devices that are only bound by Carnot efficiency.
However, the requirement of thermoelectric functionality may put tighter constraints that have been investigated for refrigerators~\cite{liu2021a} and heat engines~\cite{sobrino2026b}.
We therefore consider the weaker CTUR inequality~\eqref{EQ:cturfull} in this appendix, which can be rewritten as 
\begin{align}\label{EQ:cturfull1}
\sigma \ge 2 I_M \asinh\left(\frac{I_M}{S_M}\right)\,.
\end{align}

Using the above, we find that the efficiency for the heat engine regime is
\begin{align}
\eta &= \frac{(\mu_L-\mu_R) I_M}{I_E-\mu_R I_M} = \frac{(\beta_R-\beta_L)(\mu_L-\mu_R) I_M}{\sigma - \beta_L(\mu_L-\mu_R) I_M}\nn
&\le \frac{(\beta_R-\beta_L)(\mu_L-\mu_R) I_M}{2 I_M \asinh\left(\frac{I_M}{S_M}\right) - \beta_L(\mu_L-\mu_R) I_M}\nn
&= \frac{\eta_{\rm Ca}}{1+2 \frac{I_M \asinh\left(\frac{I_M}{S_M}\right)}{\beta_L P}}\,,
\end{align}
where in the first line we have expanded by $(\beta_R-\beta_L)$ and completed the denominator to create $\sigma$, then in the second line used the above inequality~\eqref{EQ:cturfull1}
and in the last line inserted $\eta_{\rm Ca} = 1 - \frac{\beta_R}{\beta_L}$ as well as $P=-(\mu_L-\mu_R) I_M$.
As this holds in regimes where $P\ge 0$ and we trivially have $I_M \asinh\left(\frac{I_M}{S_M}\right)\ge 0$ this bound is always tighter than the Carnot efficiency.
The bound corresponds to the dash-dotted red curve in the bottom panel of Fig.~\ref{FIG:currentsandefficiency}.

For the coefficient of performance, quite similar arguments can be used to obtain an upper bound 
\begin{align}
\kappa &= \frac{I_E-\mu_L I_M}{(\mu_L-\mu_R) I_M} = \frac{\beta_R(I_E-\mu_L I_M)}{\sigma + (\beta_L-\beta_R) (I_E-\mu_L I_M)}\nn
&\le \frac{\kappa_{\rm Ca}}{1+\frac{2 I_M \asinh\left(\frac{I_M}{S_M}\right)}{(\beta_L-\beta_R) (I_E-\mu_L I_M)}}\,,
\end{align}
where in the first line we expanded by $\beta_R$ and completed the denominator to create $\sigma$, then in the second line used inequality~\eqref{EQ:cturfull1} and used 
$\kappa_{\rm Ca} = \frac{\beta_R}{\beta_L-\beta_R}$.
Analogously, since this holds for $\beta_L\ge \beta_R$ and $I_Q = I_E-\mu_L I_M \ge 0$, this upper bound is below the Carnot coefficient of performance.
The bound corresponds to the dash-dotted blue curve in the bottom panel of Fig.~\ref{FIG:currentsandefficiency}.

\end{document}